\documentclass[aps,prc,groupedaddress,showpacs,manuscript]{revtex4-1}
\usepackage{graphicx}

\usepackage{colordvi}

\begin{document}

\title{An explanation of the elliptic flow difference between proton and anti-proton from the UrQMD model with hadron potentials}

\author {Qingfeng Li$\, ^{1}$\footnote{E-mail address: liqf@hutc.zj.cn},
Yongjia Wang$\, ^{1}$,
Xiaobao Wang$\, ^{1}$,
Caiwan Shen$\, ^{1}$
}

\affiliation{
1) School of Science, Huzhou University, Huzhou 313000, P.R. China \\
\\
 }
\date{\today}

\begin{abstract}

The time evolution of both proton and anti-proton $v_2$ flows from Au+Au collisions at $\sqrt{s_{NN}}$=7.7 GeV are examined by using both pure cascade and mean-field potential versions of the UrQMD model. Due to a stronger repulsion at the early stage introduced by the repulsive potentials and hence much less annihilation probabilities, anti-protons are frozen out earlier with smaller $v_2$ values. Therefore, the experimental data of anti-proton $v_2$ as well as the flow difference between proton and anti-proton can be reasonably described with the potential version of UrQMD.

\end{abstract}



\maketitle
\section{Motivation}
In order to prove the existence and to further understand the properties of the quark-gluon plasma (QGP), the stiffness of the equation of state (EoS) of nuclear matter and the possible (order of) phase transition from the hadron gas (HG) to the QGP should be investigated thoroughly. Therefore, besides plentiful investigations on the theoretical and modelling side, on the experimental side the heavy-ion colliding energies of accelerators have been boosted
upwards from GSI Schwerionen Synchrotron (SIS, with beam energies from several hundred MeV/nucleon to several GeV/nucleon), BNL
Alternating Gradient Synchrotron (AGS, with beam energies of several GeV/nucleon), CERN Super Proton
Synchrotron (SPS, with beam energies from several tens GeV/nucleon to 1-2 hundred GeV/nucleon), BNL Relativistic Heavy Ion Collider
(RHIC, with nucleon-nucleon centre-of-mass energies $\sqrt{s_{NN}}$ up to several hundred GeV), till to the highest sofar CERN Large Hadron Collider (LHC, with $\sqrt{s_{NN}}$ up to several TeV). Especially, to draw firm conclusions about the location of the critical endpoint at finite baryon chemical potential $\mu_B$  (which have been hooked into the region in lower SPS energies based on more and more comparisons of calculations and corresponding experimental observables) and the boundary of the first order transition, a careful and systematic beam energy scan with several independent experimental programs at SPS (e.g., with the NA49, CERES, and NA50/NA60/NA61 experiments) and RHIC (with the BES experiments in the $\sqrt{s_{NN}}$ range of 7.7-62.4 GeV) is essential and ongoing \cite{Satz:2004zd,Adamova:2002ff,Odyniec:2013kna}.

In past several years, new interesting results measured by STAR collaboration for the phase I of the RHIC-BES program have been available \cite{Odyniec:2013kna}. Besides confirmations on previous measurements by collaborations at CERN-SPS, more attention has been drawn on new findings, such as the elliptic flow splitting of particles especially of protons and anti-protons seen at low BES energies \cite{Adamczyk:2013gw}. While a constituent quark number scaling of the hadronic elliptic flow indicates the existence of the QGP phase, the splitting might hint the change of its dynamical evolution or even its disappearance which has been explained by several theoretical groups \cite{Steinheimer:2012bn,Sun:2014rda,Xu:2013sta,Ivanov:2014zqa,Hatta:2015era}.

In Ref.~\cite{Steinheimer:2012bn}, with the help of the so-called Ultra-relativistic Quantum Molecular Dynamics (UrQMD) hybrid model in which the intermediate hot and dense phase is modeled with an ideal hydrodynamic evolution (called UrQMD/H), Steinheimer {\it et al} found that the difference in elliptic flow of protons and anti-protons is pronounced just after the hydrodynamical phase (due to the difference in chemical potential) while it is almost washed out by the subsequent transport phase in the cascade mode (due to the dominant annihilation process). In Ref.~\cite{Xu:2013sta}, using an updated version of a multiphase transport (AMPT) model in which the evolution of partons in phase space is modeled by a 3-flavor Nambu-Jona-Lasinio (NJL) model, Xu {\it et al} found that the splitting effect is attributed to different mean-field potentials for hadrons and anti-hadrons as well as quarks and anti-quarks.

In our previous investigations \cite{Li:2007yd,Li:2008ge,Li:2010ie} based on a mean-field potential version of the UrQMD (called UrQMD/M), a consideration of potential modifications for ``pre-formed'' particles (string fragments) from color fluxtube fragmentation and for confined particles helps much to describe several observables such as the Hanbury-Brown-Twiss interferometry (HBT) of two particles (especially the time-related HBT-puzzle), the elliptic flow (in the cascade mode calculations, it is known as a flow-puzzle), and the yields of strange baryons or anti-baryons (a puzzle related to the strangeness enhancement). Although a thorough explanation of all existing ``puzzles'' is still awaiting since a complete description of the multi-particle collision dynamics crossing a possible phase transition and/or a consistency with the first-principle lattice QCD calculations has not arrived yet, currently it is worthwhile examining integrants of EoS at different stages based on various mechanisms and transport models so that a true and necessary connection of the observable to the physical reason behind can then be clarified to the end.

The paper is arranged as follows. In the next section, the UrQMD model is introduced firstly, together with a brief introduction to the elliptic flow parameter $v_2$. In Sec. III, we show UrQMD calculations of elliptic flow $v_2$ of protons and anti-protons from Au+Au collisions at $\sqrt{s_{NN}}$=7.7 GeV. In addition, effects of time evolution and corresponding scattering and annihilation processes on $v_2$ are analyzed and discussed. For each reaction, more than 1 million events are considered to guarantee a sufficient statistical accuracy. And the program stops at $t_{\mathrm cut}=50$ fm$/c$ except where stated otherwise. Finally, a summary and outlook is given in Sec. IV.

\section{UrQMD model and collective flows}
It is known that the UrQMD model \cite{Bass:1997xw,urqmdweb1} originates from its two working parents named the quantum molecular dynamics (QMD) model \cite{Aichelin:1991xy} (whose various modified versions have been successfully used for modeling HICs at SIS and even lower energies) and the relativistic molecular dynamics (RQMD) model \cite{Sorge:1989dy} (whose following (hybrid) versions are being used to serve mainly for HICs at energies ranging from AGS, SPS, RHIC, up to LHC), respectively. Therefore, with a continuous maintenances and updates, a large number of theoretical analyses, predictions, and comparisons with data have been supplied by the UrQMD transport model for our community, over a larger range of beam energies.

With the cascade mode (called UrQMD/C, and the current version available is 2.3) as a good basis in which 55 baryon and 32 meson species and their corresponding anti-baryons and isospin-projected states are tabulated and produced via string or resonance decays, or via two-body s-channel collisions, the updates on UrQMD in recent years have two important branches. One is a ``hybrid version'' (UrQMD/H) in which the macroscopic hydrodynamical process is inserted \cite{Li:2008qm,Bleicher:2015vha}, the other is a ``mean-field potential version'' (UrQMD/M) in which the mean-field potentials are considered for both formed baryons and pre-formed hadrons in a similar way and expressed as \cite{Li:2007yd}

\begin{equation}
U(\rho_h/\rho_0)=\mu (\frac{\rho_h}{\rho_0})+\nu
(\frac{\rho_h}{\rho_0})^g, \label{den3}
\end{equation}
where $\mu$, $\nu$ and $g$ are parameters which decide the stiffness
of the EoS of nuclear matter (in this work, the same soft EoS with momentum dependence and with the incompressibility $K=314$ MeV is adopted). $\rho_0$ is the normal density and $\rho_h$ is the density for both pre-formed hadrons and formed baryons (including anti-baryons). For formed mesons, no nuclear potential is considered as before. Certainly, the quark number difference between pre-formed baryons and mesons, and the relativistic effect on relative distances and momenta of two particles have been taken into account in the corresponding potential modification process.  As stated in the previous section, both branches of model updates have showed the importance of the consideration with  strong interactions (or, EoS) in the hot and dense (new) phase.

As known as the second coefficient of the Fourier expansion of the azimuthal distribution of the emitted particles, the elliptic flow parameter $v_2$ focused in this work is defined as

\begin{equation}
v_2\equiv <\mathrm{cos}[2(\phi-\Phi_{\mathrm{RP}})]>=<\frac{p_x^2-p_y^2}{p_t^2}>.
\label{eqv2}
\end{equation}
Here $\phi$ denotes the azimuthal angle of the considered outgoing particle;
$\Phi_{\mathrm{RP}}$ is the azimuthal angle of the reaction plane which is pre-configured
in the calculations, $\Phi_{\mathrm{RP}}=0$, but has to be determined in high-energy experiments by the so called event plane (or participant plane) which is usually not perfectly aligned with the reaction plane due to fluctuations \cite{Petersen:2010cw};
$p_x$ and $p_y$ are the two components of the transverse
momentum $p_t=\sqrt{p_x^2+p_y^2}$. The angular brackets denote an
average over all considered particles from all events. At both very low ($E_b\lesssim 0.1$ GeV$/$nucleon) and high ($\gtrsim 10$ GeV$/$nucleon) beam energies and for free nucleons, the in-plane flow dominates, which leads to an positive $v_2$ value. While in between the beam energy, the ``squeeze out'' effect plays a more important role which can be explained by the consideration of the mean field potentials for confined baryons, and depicts a deep spoon-like structure \cite{Petersen:2006vm}.

Roughly speaking, in the concept for a many-body description in terms of a real time non-equilibrium field theory, both the mean-field potentials and the collisions between particles are equally important to reproduce quantitatively the experimental particle emissions from HICs at energies above multi-fragmentation ($E_b \sim 0.1$ GeV$/$nucleon) \cite{Mao:2005aa}. However, the multiplicity of emitted particle species and the dynamic evolution of the hot and dense nuclear medium (including density and momentum) together decide the complexity and richness of the excitation function of flow results \cite{Andronic:2006ra,Adamczyk:2013gw,Adamczyk:2014ipa}. Therefore, besides the stiffness of EoS, the treatment in collision term, especially on the medium modifications of (differential) cross sections with various incoming and outcoming particle types, ought to be paid more attention. At SIS energies, with the help of approaches such as the
(self-consistent) relativistic Boltzmann-Uehling-Uhlenbeck (RBUU) and the (Dirac-)Brueckner-Hartree-Fock (DBHF), it has been found that the yields and flows of particles and clusters are sensitive to the medium-modified nucleon-nucleon cross sections (NNCS) and the corresponding FOPI experimental data can be well described with a proper treatment on the density, isospin, and momentum dependence of NNCS \cite{Li:2005jy,Li:2006ez,Jiang:2007fv,Prassa:2007zw,Li:2011zzp,Wang:2013wca}.  However, due to the lack of both experimental data and theoretical progress, the collisions in HICs at higher beam energies are still with large uncertainties since {\it ad hoc} assumptions must be made for unknown hadron-hadron cross sections. For example, in the current version of UrQMD, differential angular distributions for all two-body collisions are treated similarly by the differential cross-section of free NN elastic scattering and, further, have no direct connection to the coressponding total cross sections, which is obviously questionable. It is noticed that in the AMPT calculations in Ref.~\cite{Xu:2013sta} the scattering between quarks and antiquarks is treated to be isotropic for simplicity.

\section{$v_2$ results for protons and anti-protons}

\begin{figure}[htbp]
\centering
\includegraphics[angle=0,width=0.8\textwidth]{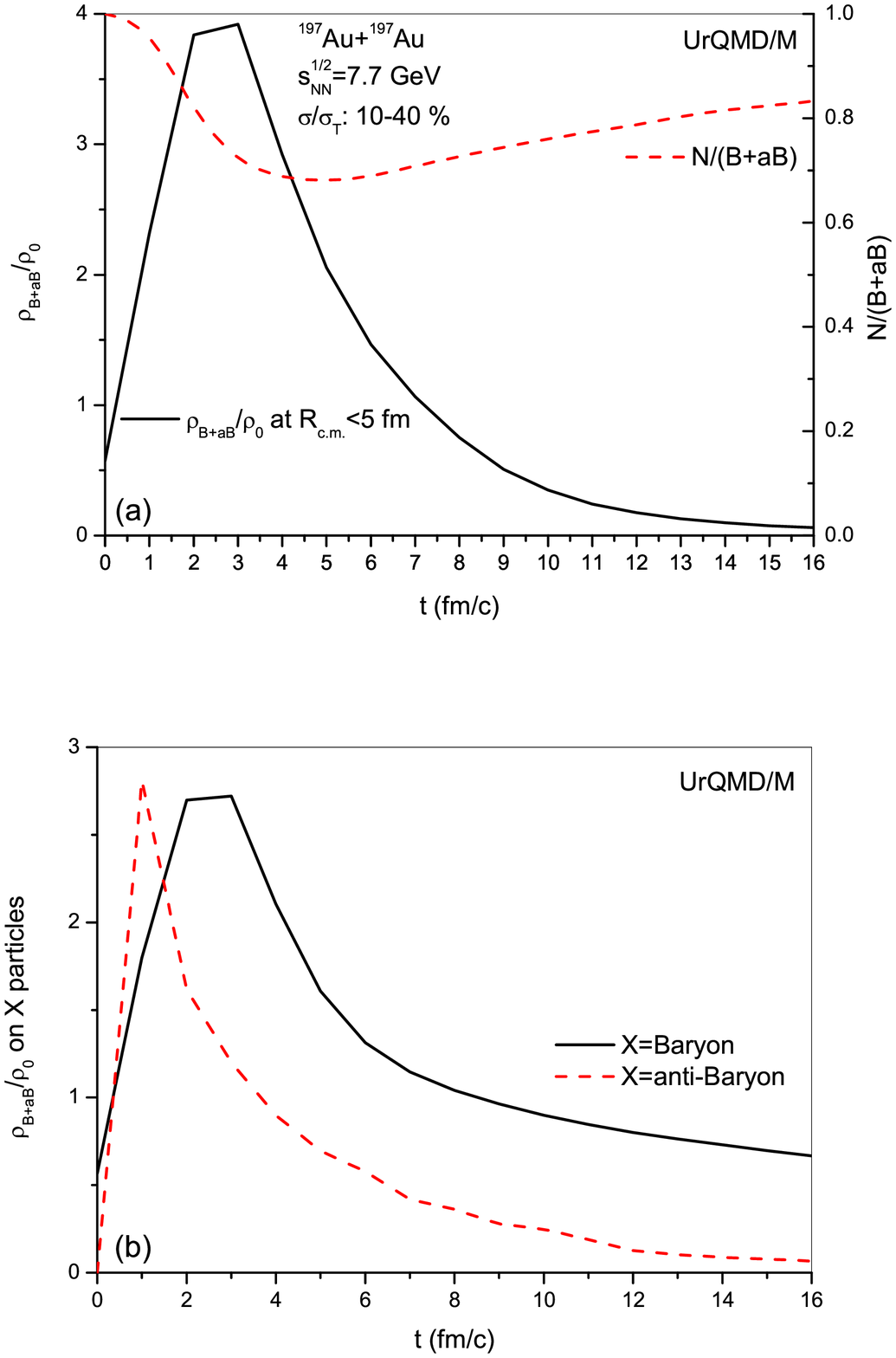}
\caption{\label{fig1} (Color online) (a): Time evolution of $\rho_{B+aB}/\rho_0$ at a central zone ($R_{c.m.}<5$ fm) (solid line, using the left axis) and the ratio of nucleons to all baryons (dashed line, using the right axis). (b): Time evolution of $\rho_{B+aB}/\rho_0$ on the positions of baryons (solid line) or anti-baryons (dashed line). Calculation with UrQMD/M is used as an example.
}
\end{figure}

First of all, the time evolution of reduced baryon density ($\rho_{B+aB}/\rho_0$) at a central zone ($R_{c.m.}<5$ fm) and on the positions of all (anti-)baryons are shown in Fig.~\ref{fig1} (a) and (b), respectively. In Fig.~\ref{fig1} (a) the maximum value of $\rho_{B+aB}/\rho_0$ (about 4) happens as early as $\sim$3 fm$/$c, and returns to normal density at about 7$\sim$8 fm$/$c. This is similar to the time evolution of baryon densities as shown in the plot (b), although the maximum value are now somewhat smaller due that the non-central collisions are in use for calculations. Because of the high baryon density happened in the centre area of each collision, a large number of new baryons are produced which evolves in the following stages with different ways from nucleons. As seen from the ratio of nucleons to all baryons (dashed line) shown in Fig.~\ref{fig1} (a), the new produced baryons can be as many as 30$\%$ of total baryons at $t=4$ fm$/$c. Since the $\sqrt{s_{NN}}$=7.7 GeV is equivalent to a projectile beam energy of $\sim 29.7$ GeV with the fixed Au target, it is found from previous calculations \cite{Li:2010ie} that a large amount of new produced baryons originate from the new mechanism---string excitation and fragmentation, which is obviously essential for anti-baryons. As seen in Fig.~\ref{fig1} (b), the (pre-formed) anti-baryons are produced at even earlier times, $\sim 1$ fm$/$c, and frozen out faster than baryons since at about $4$ fm$/$c the reduced density has already turned back to the normal one.

\begin{figure}[htbp]
\centering
\includegraphics[angle=0,width=0.8\textwidth]{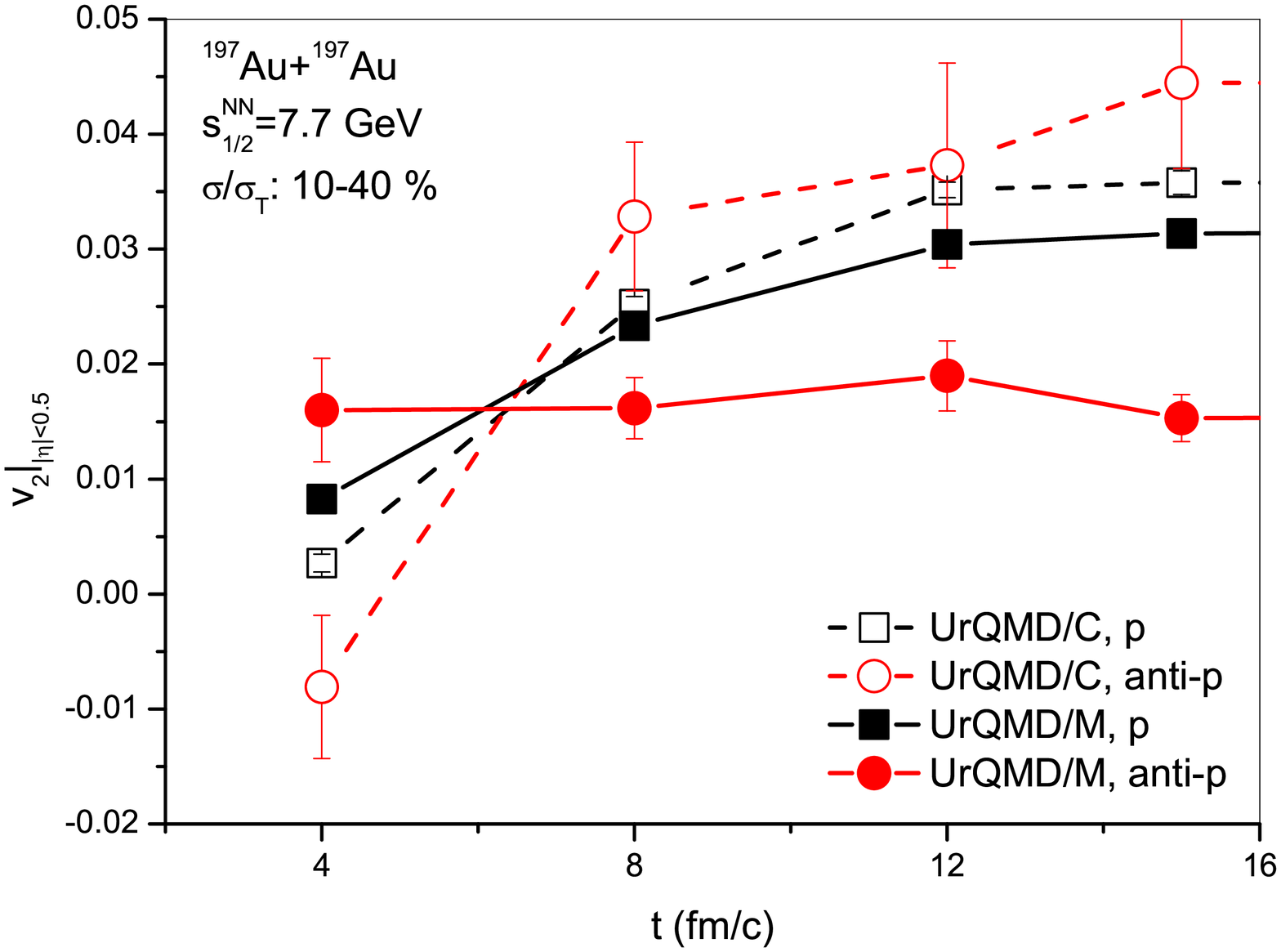}
\caption{\label{fig2} (Color online) Time evolution of $v_2$ of proton (squares) and anti-proton (circles) in the mid-rapidity region with a cut $|y|<0.5$. Calculations with a pure cascade mode (dashed lines with open symbols) are compared to results with UrQMD/M (solid lines with solid symbols).
}
\end{figure}

Due to the quick time-evolution of high baryon density (or, the difference of pressure gradient) in the colliding centre, the quick growth of $v_2$ at mid-rapidities takes place also at the early stage, which was depicted in the previous publication \cite{Petersen:2006vm} (for pions there) and is also shown in Fig.~\ref{fig2} (for protons and anti-protons here). The UrQMD/C calculations (dashed lines with open symbols) are compared to the results with UrQMD/M (solid lines with solid symbols). A pseudo-rapidity interval of $|\eta|<0.5$ is used for all particles. In the cascade mode, $v_2$ values of both protons and anti-protons increase obviously in the time duration $4-12$ fm$/$c. At $t=4$ fm$/$c, due that the newly created pre-formed particles from string fragmentation are treated to be free-streaming, the $v_2$ of anti-protons is seen even smaller than that of protons. With increasing time, and with the strong annihilation effect on anti-protons mainly at low momenta, the $v_2$ of anti-protons increases more quickly than that of protons and finally, even slightly larger than the latter. With the consideration of mean-field potentials, $v_2$ values of both particles are seen to be enhanced at the early time $t=4$ fm$/$c due to a stronger early pressure. With the increase of time from 4 to 8 fm$/$c, the proton $v_2$ increases further due to both the repulsive potentials and a large number of two-body collisions. However, during the same period, the anti-proton $v_2$ keeps unchanged. It is understandable since most of anti-protons have been pushed out of the fireball by the strong repulsive potential and survived without a further annihilation process as well as potential modifications, which has been seen in a previous calculation shown in Ref.~\cite{Li:2010ie}. If the potentials are only considered for formed baryons, however, the annihilation effect on anti-protons is still strong due to the lack of pressure at the earlier stage, and the anti-proton $v_2$ will follow with the increase of proton $v_2$. This is similar to calculation results with UrQMD/H in Ref.~\cite{Steinheimer:2012bn}. From 8 to 12 fm$/$c, the central density falls to the sub-normal densities, which has caused a bulky attractive-potential effect. Meanwhile, owing to a stronger repulsion at the earlier stage so as to an earlier freeze-out, the total collision number becomes also smaller. Therefore, the proton $v_2$ with UrQMD/M is even smaller than that with UrQMD/C. It should be noticed that at higher beam energies (such as at $\sqrt{s_{NN}}$=200 GeV) and for more central collisions, the strong repulsive potential could even drive up the $v_2$ value, as seen in Ref.~\cite{Li:2007yd}.

More interestingly, hereafter, a $v_2$ flow difference between proton and anti-proton is seen in UrQMD/M calculation results. And, from above analyses, this originates from a consideration of the mean-field potential mainly on pre-formed hadrons. This potential effect is similar to the explanation by the AMPT calculations in which a NJL model is inserted for the evolution of partons \cite{Xu:2013sta}. But, they are different in details. In Ref.~\cite{Xu:2013sta}, it is understood from the repulsive and attractive effects of the partonic vector potential on quarks and antiquarks, respectively. Here a repulsive early pressure on pre-formed hadrons affords enough effect on driving down anti-proton $v_2$ value because it freezes out earlier and survives from the annihilation process. \footnote{since at low collision energies protons are much more
abundant than anti-protons, the annihilation only affects the anti-protons.} Certainly, an attractive potential also exists in our model, but it appears at the late stage when the baryon density becomes low.

\begin{figure}[htbp]
\centering
\includegraphics[angle=0,width=0.8\textwidth]{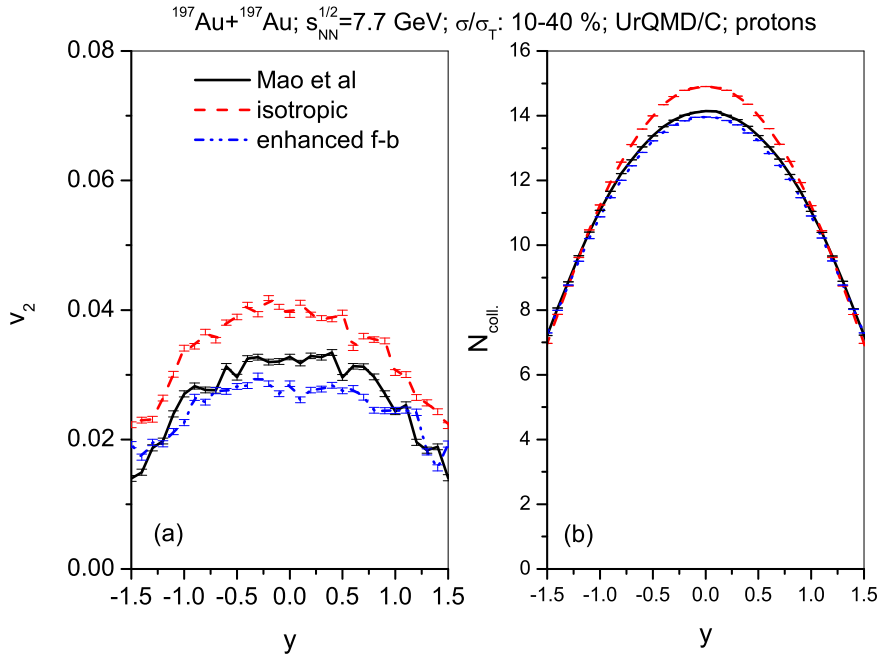}
\caption{\label{fig3} (Color online) (a): Rapidity $y$ dependence of the proton $v_2$ with three types of angular distribution for all two-body processes (see text). (b): Corresponding rapidity distributions of the collision numbers.
}
\end{figure}

Hence, on an equal footing, this finding stresses the importance of the respective scattering and annihilation processes. Besides anti-proton $v_2$, the proton $v_2$ should be also heavily influenced by the treatment in scattering process since the current calculated $v_2$ with UrQMD/M is only about $2/3$ of experimental data. Beside the treatments of the nuclear medium modifications on cross sections \cite{Li:2005jy,Li:2006ez,Jiang:2007fv,Prassa:2007zw,Li:2011zzp,Wang:2013wca,Li:2008gp}, the differential cross sections of each channel should be paid much more attention since it is always simply treated in the running transport models and usually has no any (or very loose) connection with final total cross sections. Fig.~\ref{fig3} (a) shows the rapidity $y$ dependence of the proton $v_2$ with three types of angular distribution for all two-body processes. The solid line represents results with the default one used in UrQMD, which was developed by Mao {\it et al} originally for free NN elastic scattering (please see Refs.~\cite{Bass:1997xw,Mao:2005aa} and references given therein). It is known from Mao's calculations that the differential NN elastic cross section in free space is forward-backward peaked, especially with the increase of incident energy. However, taking the nuclear density into account, the angular distribution might be more isotropic or more forward-backward peaked, depending on theoretical hypotheses on various reaction channels \cite{Mao:1994zza,Mao:1997gr,AbdelWaged:2004wk,White:2014nia}.  The dashed line in Fig.~\ref{fig3} (a) depicts the result using an isotropic angular distribution, but for the dash-dot-dotted line we artificially let Mao's distribution be two times more forward-backward peaked. It is clearly seen that a more isotropic distribution causes a larger $v_2$, especially at mid-rapidities, which is reasonable due to a more transverse emission. Correspondingly, a larger number of collisions is seen, which is shown in Fig.~\ref{fig3} (b). If we check the relationship between $v_2$ values in Fig.~\ref{fig3} (a) and collision numbers in Fig.~\ref{fig3} (b), it is found that the increase of flow follows monotonously with the increase of the total collision number determined by cross sections. Further, the $v_2$ value seems to be more sensitive to the treatment in differential cross sections. This phenomenon can be explained with an azimuthal anisotropic escape mechanism. It is noticed from Ref.~\cite{He:2015hfa} that, using the string melting version of the AMPT model for HICs at $\sqrt{s_{NN}}$=7.7 GeV, the majority of flows is found to come from the anisotropic escape probabilities of partons, which is in line with our UrQMD findings.

\begin{figure}[htbp]
\centering
\includegraphics[angle=0,width=0.8\textwidth]{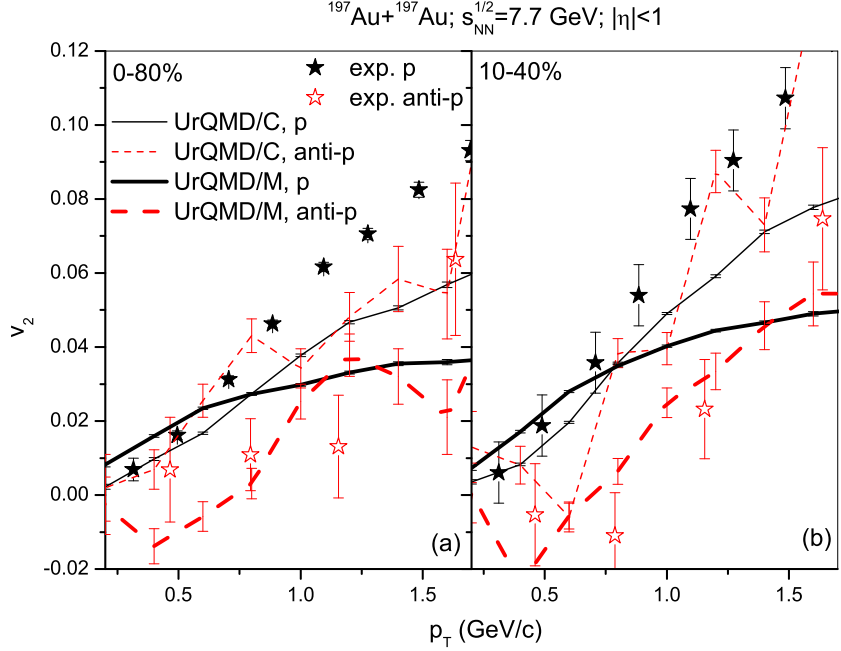}
\caption{\label{fig4} (Color online) Transverse momentum $p_t$ dependence of proton (solid black lines) and anti-proton (dashed red lines) $v_2$ from 0$\%$-80$\%$ (in plot (a)) and 10$\%$-40$\%$ (in plot (b)) central Au+Au collisions at $\sqrt{s_{NN}}$=7.7 GeV, respectively. The STAR data are taken from Ref.~\cite{Adamczyk:2013gw}.
}
\end{figure}

Finally, Fig.~\ref{fig4} depicts the transverse momentum $p_t$ dependence of proton (solid black lines) and anti-proton (dashed red lines) $v_2$ from 0$\%$-80$\%$ (in plot (a)) and 10$\%$-40$\%$ (in plot (b)) central Au+Au collisions at $\sqrt{s_{NN}}$=7.7 GeV, respectively. At both centralities, both UrQMD/C (thin lines) and UrQMD/M (thick lines) calculations are compared with STAR data \cite{Adamczyk:2013gw} of protons (solid black star symbols) and anti-protons (open red star symbols). For comparison, a pseudorapidity cut $|\eta|<1$ is used. As seen above, an obvious flow $v_2$ difference between protons and anti-protons is seen in the selected $p_t$ window from UrQMD/M, while it is not reproduced by UrQMD/C calculations. Meanwhile, within UrQMD/M calculations, the anti-proton $v_2$ becomes smaller and can describe the data fairly well. Taking the factors discussed above into consideration, we realize that the difference of pressure gradient caused by the initial geometric structure and the evolution of baryon density leads to the quick increase of $v_2$ at the early stage. With the consideration of the repulsive potentials, the total collision number will be decreased, and results in an earlier freeze-out of particles. This leads to the increase (decrease) of proton $v_2$ at low (high) transverse momenta. For anti-protons, due to the much earlier freeze-out and much less annihilation probabilities, most of them are survived and the $v_2$ becomes much smaller. To quantitatively improve the $p_t$ dependence of proton $v_2$, besides a more careful consideration on the stiffness of EoS, one needs to comprehensively comb the treatments of the initial condition \cite{Zhao:2014lka} and collision cross sections (including differential ones) of particles (including pre-formed and formed ones) used in UrQMD. In addition, in Ref.~\cite{Soff:1999et}, it was found that, considering the multi-gluon exchange process, an effectively larger string tension of $\kappa=3$ GeV$/$fm (the default one in UrQMD is 1 GeV$/$fm in a freely colored environment), accordingly shorter formation times of the strings, described much better the experimentally observed high values of hyperon yields, as well as the elliptic flow values especially at about $p_t>1$ GeV$/$c. However, it should be noticed that the increase of string tension $\kappa$ value leads to a larger slope value of the directed flow $v_1$ ($=<p_x/p_t>$) at mid-rapidity as well, which is obviously not in agreement with $v_1$ experimental data as discussed in Refs.~\cite{Adamczyk:2014ipa,Steinheimer:2014pfa}. Therefore, a more systematic consideration of the collision process in UrQMD is urgently needed, which is in progress.

\section{Summary and Outlook}
In summary, we use both pure cascade and mean-field potential versions of the UrQMD model to calculate the time evolution of both proton and anti-proton $v_2$ flows from Au+Au collisions at $\sqrt{s_{NN}}$=7.7 GeV. It is found that due to a stronger repulsion at the early stage introduced by the repulsive potentials and hence much less annihilation probabilities, the experimental data of anti-proton $v_2$ as well as the flow difference between proton and anti-proton can be reasonably described with the potential version of UrQMD. In order to reproduce the transverse momentum $p_t$ dependence of $v_2$ for protons, more theoretical endeavors are worthwhile shedding light on some rather unclear ingredients of the whole dynamical process, such as the stiffness of EoS for both HG and QGP phases, and the modifications of two-body scattering (differential) cross sections.

\begin{acknowledgements}
We thank Profs. M. Bleicher and F.Q. Wang for useful discussions and acknowledge support by the computing server C3S2 in Huzhou
University. The work is supported in part by the National
Natural Science Foundation of China (Nos. 11375062, 11275068), the project sponsored by SRF for ROCS, SEM, and the Doctoral Scientific Research Foundation (No. 11447109).
\end{acknowledgements}

\end{document}